\documentstyle[preprint,aps]{revtex}

\begin{document}
\draft
\tighten
\title{Coulomb blockade oscillations of conductance in the regime of
  strong tunneling}

\author{A.~Furusaki\cite{AF} and K.~A.~Matveev}

\address{Department of Physics, Massachusetts Institute of Technology,
  Cambridge, MA 02139}

\date{\today}
\maketitle

\begin{abstract}
  We study the transport through a quantum dot coupled to two leads by
  single-mode point contacts.  The linear conductance is calculated
  analytically as a function of a gate voltage and temperature $T$ in the
  case when transmission coefficients of the contacts are close to unity.
  As a function of the gate voltage, the conductance shows Coulomb
  blockade oscillations. At low temperatures, the off-resonance
  conductance vanishes as $T^2$, in agreement with the theory of inelastic
  co-tunneling.  Near a resonance, the low-energy physics is governed by a
  multi-channel Kondo fixed point.
\end{abstract}

  The Coulomb blockade of tunneling has recently become a subject of
  intensive studies \cite{AL}.  It is usually observed by measuring the
  conductance of a system of two bulk electrodes connected by tunnel
  junctions to a small conducting island. Tunneling of an electron into
  the island is accompanied by the increase of the energy of the system by
  $E_C=e^2/2C$, where $C$ is the capacitance of the island. At low
  temperatures $T\ll E_C$ this leads to the suppression of tunneling. This
  phenomenon is due to the discreteness of charge in the island, and can
  be suppressed by tuning a gate voltage to the point where the energies
  of the states with $n$ and $n+1$ electrons in the island are equal. At
  these points the energy gap related to the charging energy vanishes, and
  one observes peaks in conductance as a function of the gate voltage.

  A popular realization\cite{Kastner} of the conducting island is a
  quantum dot, Fig.~\ref{fig:1}, created artificially in a two-dimensional
  electron gas (2DEG), and connected to the large areas of 2DEG (the
  leads) by quantum point contacts (QPC).  In such a system, the
  transmission coefficient ${\cal T}$ of a QPC connecting the dot and an
  external lead can be controlled by changing a gate voltage.  This opens
  a possibility for studying the Coulomb blockade effect in the
  strong-tunneling regime of ${\cal T}\to1$, where one naively expects
  that the Coulomb blockade may not be observed because the number of
  electrons in a quantum dot is no longer well defined.  Recent
  experiments indicate, however, that even when ${\cal T}$ is close to 1
  the conductance shows periodic oscillations although the peaks are not
  well separated \cite{van,Pasquier}.  Most of theoretical work has been
  concentrated on the case of weak tunneling (${\cal T}\ll1$), and only
  the equilibrium thermodynamic quantities have been discussed in the
  strong-tunneling limit \cite{Matveev1}.  The aim of this paper is to
  develop a theory of the transport through a quantum dot in the
  strong-tunneling regime.  We show that in the low-energy limit the
  conductance is renormalized to zero off resonance, and to $\sim e^2/h$
  on resonance, yielding clear Coulomb blockade peaks even for the
  strong-tunneling case.  We also derive analytic expressions for the
  conductance in some interesting limiting cases.

In the weak-tunneling limit there are two different mechanisms of
low-temperature conductance. One contribution is due to real transitions
of electrons between the leads and the dot\cite{Glazman}. If the gate
voltage $V_g$ is not equal exactly to the nearest resonant value
$V_g^{(n)}$, an energy cost $\Delta E\propto V_g-V_g^{(n)}$ is associated
with such a transition. Since only an exponentially small fraction of
electrons have energy $\Delta E$ at low temperatures, the conductance
decays exponentially away from the peak:
\begin{equation}
G=\frac{1}{2}\frac{G_LG_R}{G_L+G_R}
  \frac{\Delta E/T}{\sinh(\Delta E/T)}.
\label{GS}
\end{equation}
Here $G_L$ and $G_R$ are the conductances of the left and right QPCs.  At
very low temperatures another mechanism of transport through the dot
dominates \cite{AN}. This mechanism, commonly referred to as the inelastic
co-tunneling, corresponds to the second-order tunneling processes. At the
first step an electron tunnels from the left lead to a virtual state in
the dot, and the system acquires charging energy $\sim\Delta E$. At the
second step, another electron tunnels from the dot to the right lead, thus
finishing the process of the charge transfer and restoring the initial
charge of the dot.  In any such process, the original electron decays into
three quasiparticles (an electron in the right lead and an electron-hole
pair in the dot). Similarly to the problem of decay of a
quasiparticle in Fermi liquid, this means that the tunneling rate is
proportional to $T^2$ in the low-temperature limit,
\begin{equation}
G=\frac{G_LG_R}{6G_0}\left(\frac{T}{\Delta E}\right)^2,
\quad
T\ll\Delta E,
\label{cotunneling}
\end{equation}
where $G_0=e^2/h$. In the case of weak tunneling, $G_{L,R}\ll G_0$, the
co-tunneling mechanism gives only a small correction to the peak value
given by Eq.~(\ref{GS}). However, it dominates away from peaks at $T\to0$.

At low temperatures and at the gate voltage near the peak position, i.e.,
at $T, \Delta E\ll e^2/2C$, the tunneling matrix elements are renormalized
and grow logarithmically\cite{GM}.  This can be seen in the following way
\cite{Matveev2}.  On resonance, the state with $n$ electrons in the
quantum dot and that with $n+1$ electrons have the same electrostatic
energy, i.e., $\Delta E=0$.  These two states can be regarded as up and
down states of a fictitious impurity ``spin'' $S=\frac{1}{2}$, and we may
discard all other states. If we also ascribe up (down) ``spin'' to the
electrons in the leads (dot), then each tunneling event flips ``spins'' of
both the tunneling electron and the impurity.  Thus, the tunneling
Hamiltonian can be interpreted as ``spin''-flip scattering on an impurity.
Hence the tunneling problem is mapped to an anisotropic multi-channel
Kondo problem, in which the number of the channels (flavors) is equal to
the total number of 1D modes in all the QPCs.

In the leading logarithmic approximation \cite{GM,Matveev2} one can
substitute into Eq.\ (\ref{GS}) the renormalized $\tilde G_L$ and $\tilde
G_R$ obtained from the scaling equations for the anisotropic Kondo model,
\begin{equation}
\tilde G_{L(R)}=\frac{G_{L(R)}}
         {\cos\left[\sqrt{G_{L(R)}/2\pi^2G_0}
                                 \ln(E_C/T)\right]}.
\label{renorm-G}
\end{equation}
If $G_L$ and $G_R$ are equal, they grow together under the
renormalization, and the peak conductance will be of the order of $G_0$ at
low temperature\cite{renorm-limit}.  When $G_L$ is initially smaller than
$G_R$, however, they first increase together according to
(\ref{renorm-G}), but then $G_L$ starts to decrease to zero whereas $G_R$
keeps increasing to $G_0$.  This is because the fixed point of the
multi-channel Kondo problem is unstable against a perturbation breaking
the flavor symmetry \cite{NB}.  Thus the total conductance in this case
shows a nonmonotonic temperature dependence and goes to zero in the
low-temperature limit.

Below we concentrate on the case of strong tunneling, and
find much stronger renormalizations of conductance than the ones given by
Eq.~(\ref{renorm-G}).
The system we study is a quantum dot connected to two external leads by
single-mode QPCs.  We consider the low-temperature case $ T\ll E_C$,
but assume that the level spacing in the dot is much smaller than the
temperature.  The latter assumption is usually satisfied for reasonably
large quantum dots. It means that the phase coherence in transport of
electrons from one QPC to the other is destroyed by thermal fluctuations,
and one can neglect the corresponding processes of elastic
co-tunneling\cite{AN}.  Since the transport through a single-mode QPC is
essentially one dimensional (1D), we may introduce for each QPC an
effective 1D model with linearized dispersion relation \cite{note2}.  We
further assume that the Coulomb repulsion can be described by the charging
energy $Q^2/2C$ because of good screening in the quantum dot, and use
point-like backward-scattering potential to model reflection at the
QPCs.  The two 1D systems are coupled by the charging energy.  The
effective Hamiltonian is
\begin{eqnarray}
H&=&
v_F\int^\infty_{-\infty}\!\! dx\sum_{j=L,R}
            \sum_{\sigma=\uparrow,\downarrow}\left[
\psi^\dagger_{j,1,\sigma}(x)(i\partial_x-k_F)\psi_{j,1,\sigma}(x)
-\psi^\dagger_{j,2,\sigma}(x)(i\partial_x+k_F)\psi_{j,2,\sigma}(x)
\right]+\frac{(Q-eN)^2}{2C}
\nonumber\\
&&
+v_F\sum_{\sigma=\uparrow,\downarrow}\left\{
    |r_L|\left[\psi^\dagger_{L,1,\sigma}(0)\psi_{L,2,\sigma}(0)
           +\psi^\dagger_{L,2,\sigma}(0)\psi_{L,1,\sigma}(0)\right]\right.
\nonumber\\
 &&\hspace{5em}
\left.+|r_R|\left[\psi^\dagger_{R,1,\sigma}(0)\psi_{R,2,\sigma}(0)
           +\psi^\dagger_{R,2,\sigma}(0)\psi_{R,1,\sigma}(0)\right]
               \right\},
\label{Hamiltonian1}
\end{eqnarray}
  where $\psi_{L,1(2),\sigma}(x)$ is the field operator of a left-going
  (right-going) electron near the left QPC, $\psi_{R,1(2),\sigma}(x)$ is
  that of an electron near the right QPC; dimensionless parameter $N$ is
  proportional to the gate voltage. The charge in the quantum
  dot is given by
\begin{eqnarray}
Q= e\int_{0}^{\infty}
\sum_{d=1,2}\sum_{\sigma=\uparrow,\downarrow}
&&\left[:\!\psi^\dagger_{L,d,\sigma}(x)\psi_{L,d,\sigma}(x)\!\!:\right.
\nonumber\\
&&\left.+
:\!\psi^\dagger_{R,d,\sigma}(-x)\psi_{R,d,\sigma}(-x)\!\!:\right]dx.
\label{Q}
\end{eqnarray}

We first consider the case of spinless fermions, which turns out to be
equivalent to the two-channel Kondo problem.  This case may be realized
experimentally by applying a magnetic field parallel to the 2DEG to allow
only spin-up electrons to transmit through the QPCs.  Following the
standard procedure \cite{Sol}, we bosonize the Hamiltonian
(\ref{Hamiltonian1}):
\begin{eqnarray}
H &=&
\frac{v_F}{2}\int^\infty_{-\infty}\!\! dx\sum_{j=L,R}\left(
\frac{1}{\pi}\left[\partial_x\phi_{j}(x)\right]^2
+\pi\left[\Pi_{j}(x)\right]^2\right)
\nonumber\\
&&+\frac{E_C}{\pi^2}[\phi_{R}(0)-\phi_{L}(0)-\pi N]^2
\nonumber\\
&&+\frac{D|r_L|}{\pi}\cos[2\phi_{L}(0)]
+\frac{D|r_R|}{\pi}\cos[2\phi_{R}(0)],
\label{Hamiltonian2}
\end{eqnarray}
where $\phi_{j}(x)$ is a phase field describing charge density
fluctuations, $[\phi_{j}(x),\Pi_{k}(y)]=i\delta_{j,k}\delta(x-y)$, and
$D$ is the high-energy cutoff (bandwidth).  We assume that the reflection
amplitudes are small, $|r_{L(R)}|\ll 1$.

The current through the quantum dot is
$I=(e/2\pi)\partial_t[\phi_{L}(0)+\phi_{R}(0)]$, and the
conductance $G$ is calculated using the Kubo formula.
Up to the second order in $r_{L(R)}$ we obtain
\begin{eqnarray}
G&=&\frac{e^2}{2h}\!\left(
1-\frac{\pi\Gamma_0(N)}{4 T}
\right)\!,
\label{G1}\\
\Gamma_0(N)&=&\frac{2\gamma
E_C}{\pi^2}\left[|r_L|^2+|r_R|^2+2|r_L||r_R|\cos(2\pi N)\right],\nonumber
\end{eqnarray}
where $\gamma=e^{\bf C}$, with ${\bf C}=0.5772\ldots$ being the Euler's
constant.  We see that the second term diverges at low
temperature\cite{Flensberg}, unless $|r_L|=|r_R|$, and $N$ is a
half-integer \cite{Furusaki}. This indicates that the higher-order terms
in $|r|$ should be taken into account in a proper way.  Since the charging
mode $\phi_{R}-\phi_{L}$ is massive due to the charging energy, we may
integrate it out to obtain an effective Hamiltonian for the current mode,
$\phi_{L}+\phi_{R}$.  The resulting Hamiltonian is equivalent to that
of a single impurity in the $g=\frac{1}{2}$ Luttinger
liquid\cite{Kane,g=1/2}, which can be solved exactly \cite{Kane,Guinea}.
We use an alternative exact solution \cite{Matveev1} and fermionize the
problem to the following quadratic form,
\begin{equation}
H=\int\! dk\left[
\xi_kc^\dagger_kc_k
-\left(\lambda c^\dagger_k(c+c^\dagger)+\lambda^*(c+c^\dagger)c_k\right)
\right],
\label{Hamiltonian4}
\end{equation}
where $c_k$ and $c$ are fermions, $\xi_k=v_Fk$, and parameter
$\lambda=(\gamma v_F E_C/2\pi^3)^{1/2} (|r_L|e^{-i\pi N}+|r_R|e^{i\pi
  N})$.  The current is now given by $I=ev_F\int
:\!c^\dagger_{k_1}c_{k_2}\!\!:dk_1dk_2$.  After some algebra we get the
conductance (Fig.~2)
\begin{equation}
G=\frac{e^2}{2h}\left[
1-\int^\infty_{-\infty}dE
\left(-\frac{d f}{dE}\right)
\frac{\Gamma^2_0(N)}{E^2+\Gamma^2_0(N)}
\right],
\label{G2}
\end{equation}
where $f(E)=(e^{E/T}+1)^{-1}$.  We see that even for half-integer $N$,
i.e., on resonance, the conductance vanishes as $T^2$ if $|r_L|$ and
$|r_R|$ are not equal.  In the off-resonance case the conductance also
vanishes as $T^2$, in agreement with the result (\ref{cotunneling}) of the
inelastic co-tunneling theory.  On resonance, at $|r_L|=|r_R|$, the
conductance equals $e^2/2h$. As expected, the conductance (\ref{G2})
coincides with the one for a single impurity in the $g=\frac12$ Luttinger
liquid\cite{Kane}.

We next take into account the spins of electrons.  The bosonized form of
the Hamiltonian (\ref{Hamiltonian1}) is
\begin{eqnarray}
H&=&
\frac{v_F}{2}\int\! dx\!\sum_{j=L,R}\!\left(
\frac{1}{\pi}[\partial_x\phi_{j,c}(x)]^2+\pi[\Pi_{j,c}(x)]^2
+\frac{1}{\pi}[\partial_x\phi_{j,s}(x)]^2+\pi[\Pi_{j,s}(x)]^2\!
\right)\!
\nonumber\\
&&
+\frac{2E_C}{\pi^2}\!
\left(\phi_{R,c}(0)-\phi_{L,c}(0)-\frac{\pi}{\sqrt{2}}N\right)^2
\nonumber\\
&&
+\frac{2D|r_L|}{\pi}
\cos\left(\sqrt{2}\phi_{L,c}(0)\right)
\cos\left(\sqrt{2}\phi_{L,s}(0)\right)
+\frac{2D|r_R|}{\pi}
\cos\left(\sqrt{2}\phi_{R,c}(0)\right)
\cos\left(\sqrt{2}\phi_{R,s}(0)\right),
\label{Hamiltonian5}
\end{eqnarray}
  where $\phi_{j,c}(x)$ and $\phi_{j,s}(x)$ are the phase fields for the
  charge and spin density fluctuations.  The electric current is given by
  $I=(e/\sqrt{2}\pi)\partial_t[\phi_{L,c}(0)+\phi_{R,c}(0)]$.  Up to the
  order $|r|^2$ the conductance is
\begin{equation}
G=
\frac{e^2}{h}\left(1-
\frac{2\Gamma(\frac{3}{4})}{\Gamma(\frac{1}{4})}
\sqrt{\frac{\gamma E_C}{\pi T}}
\left(|r_L|^2+|r_R|^2\right)
\right).
\label{G3}
\end{equation}
The term proportional to $|r|^2$ has no dependence on $N$ and diverges as
$1/\sqrt{T}$ at $T\to0$, indicating that $|r_L|=|r_R|=0$ is an unstable
fixed point even on resonance, in contrast to the spinless case discussed
above.  If $|r_L|=|r_R|>0$, and $N$ is a half-integer, then the system
will be renormalized toward the fixed point of the four-channel Kondo
problem.  In fact, after a series of transformations we could map the
Hamiltonian to the one which appeared in the study of the four-channel
Kondo problem \cite{Fabrizio}.  Unfortunately, in this case we cannot sum
up analytically all the higher order divergent terms.

On the other hand, we can still find the low-temperature asymptotics of
the conductance in a realistic case when the reflection amplitudes $r_L$
and $r_R$ are not precisely equal. In this case, as we already mentioned
above, one can use mapping to the multi-channel Kondo model to identify
the stable fixed point of the problem. Since the channel anisotropy is a
relevant perturbation, in the low-energy limit the larger of the two
reflection amplitudes, say $|r_L|$, is renormalized to unity
(weak-tunneling limit), whereas $r_R$ is renormalized to zero.  Thus it is
meaningful to study the case where the transmission amplitude $t_L$ of the
left QPC and the reflection amplitude $r_R$ of the right QPC are very
small.  This limit can also be easily realized experimentally by tuning
voltages on the gates controlling the QPCs, Fig.~\ref{fig:1}.

We will calculate the conductance in the lowest order in the tunneling
probability through the left barrier. The problem is therefore to
calculate the renormalization of the tunneling density of states for the
left-lead electrons. The renormalization is due to the electrostatic
coupling between the two 1D electron systems describing the two QPCs, see
Eq.~(\ref{Hamiltonian1}). When an electron tunnels through the left
barrier, the number of particles in the dot changes by 1, which means that
the system of electrons of the right QPC is no longer in the ground state.
This leads to the suppression of tunneling due to the
orthogonality catastrophe. The resulting tunneling conductance is
\cite{Flensberg}
\begin{equation}
G=G_L\int_{-\infty}^\infty d\omega
  \frac{\frac{\omega}{T}e^{\omega/T}-e^{\omega/T} +1}{(e^{\omega/T}-1)^2}
  {\rm Re}\, K(\omega),
\label{cond-asymmetric}
\end{equation}
where $G_L\ll G_0$ is the conductance of the left barrier, and $K$ is the
correlator of the operators shifting the potential for electrons in the
right QPC: $N\to N+1$.

At $r_R=0$ the Hamiltonian for the right QPC is quadratic, and the
correlator $K(\omega)$ can be found exactly. In this case only the charge
mode $\phi_{R,c}$ is affected by the shift operator, and $K(\omega)=
K_c(\omega)= (\pi^2/\gamma E_C)f(-\omega)$.
The resulting conductance is
\begin{equation}
G=\frac{\pi^3T}{8\gamma E_C}G_L, \quad r_R=0.
\label{G4}
\end{equation}

An attempt to treat the reflection amplitude $r_R$ perturbatively leads to
the correction, similar to the second term in Eq.~(\ref{G1}), which
diverges at $T\to0$,\cite{Flensberg}. To get the non-perturbative result,
one can first notice that the system of electrons in the right QPC has
only two channels. Thus one can use the same technique\cite{Matveev1} as
in the derivation of Eq.~(\ref{G2}). First, we integrate out the charge
modes, and then the remaining Hamiltonian is fermionized to the form
(\ref{Hamiltonian4}). Clearly, the tunneling of an electron through the
left barrier affects the fermionized spin modes by shifting $N\to N+1$ and
thus changing the sign of $\lambda$ in Eq.~(\ref{Hamiltonian4}). For the
quadratic Hamiltonian (\ref{Hamiltonian4}) the calculation of the
corresponding contribution
$K_s(\omega)=4f(-\omega)\Gamma_R/(\Gamma_R^2+\omega^2)$ to the correlator
$K(t)=K_c(t)K_s(t)$ is straightforward; here
$\Gamma_R=(8\gamma/\pi^2)E_C|r_R|^2\cos^2(\pi N)$. As a result,
we get
\begin{equation}
G=\frac{\pi^2TG_L}{8\gamma E_C}
\int^\infty_{-\infty}dE
\frac{\Gamma_R}{E^2+\Gamma^2_R}
\frac{1+(E/\pi T)^2}{\cosh^2(E/2T)}.
\label{G5}
\end{equation}
When $N$ is not a half-integer, i.e., off resonance, the conductance
vanishes as $T^2$, which is again in agreement with the theory of
inelastic co-tunneling. On the other hand, on resonance the conductance
has a linear temperature dependence (\ref{G4}).

In summary, we have developed a theory for the inelastic transport through
a quantum dot via single-mode quantum point contacts in the
strong-tunneling regime.  We have obtained analytic expressions for the
peaks in linear conductance.  At any tunneling strength, between the peaks
the conductance vanishes at low temperatures as $T^2$, in agreement with
the theory of inelastic co-tunneling developed for the weak-tunneling case
\cite{AN}.  In the case of symmetric barriers the resonant value of
conductance is $\sim e^2/h$.

We are grateful to L.~I.~Glazman and P.~A.~Lee for helpful discussions.
The work was sponsored by Joint Services Electronics Program Contract
DAAL03-92-C-0001.

\begin{figure}
\caption{Schematic view of a quantum dot connected to two bulk 2D
  electrodes. The dot is formed by applying negative voltage to the gates
  (shaded). Solid line shows the boundary of the 2D electron gas (2DEG).
  Electrostatic conditions in the dot are controlled by the voltage
  applied to the central gates. Voltage $V_{l,r}$ applied to the auxiliary
  gates controls the transmission probability through the left and right
  constrictions.}
\label{fig:1}
\end{figure}

\begin{figure}
\caption{Conductance (\protect\ref{G2}) as a function of the dimensionless gate
  voltage $N$ for the symmetric case, $|r_L|=|r_R|=0.3$.  The three curves are
  calculated for $E_C/T=1,10$, and $100$.}
\label{fig:2}
\end{figure}


\begin{references}

\bibitem[*]{AF} On leave of absence from Department of Applied Physics,
  University of Tokyo, Hongo, Tokyo 113, Japan.

\bibitem{AL} For a review, see D.~V.~Averin and K.~K.~Likharev, in {\it
    Mesoscopic Phenomena in Solids}, edited by B.~Altshuler, P.~A.~Lee,
  and R.~A.~Webb (Elsevier, Amsterdam, 1991).

\bibitem{Kastner} M.~A.~Kastner, Rev.\ Mod.\ Phys.\ {\bf 64}, 849 (1992).

\bibitem{van} N.~C.~van der Vaart {\it et al.}, Physica B {\bf 189}, 99
  (1993).

\bibitem{Pasquier} C.~Pasquier {\it et al.}, Phys.\ Rev.\ Lett.\ {\bf 70},
  69 (1993).

\bibitem{Matveev1} K.~A.~Matveev, Phys.\ Rev.\ B {\bf 51}, 1743 (1995).

\bibitem{Glazman} L.~I.~Glazman and R.~I.~Shekhter, J.\ Phys.\ Conden.\
  Matter {\bf 1}, 5811 (1989).

\bibitem{AN} D.~V.~Averin and Yu.~V.~Nazarov, Phys.\ Rev.\ Lett.\ {\bf
    65}, 2446 (1990); D.~V. Averin and A.~A. Odintsov, Phys. Lett. A {\bf
    140}, 251 (1989).

\bibitem{GM} L.\ I.\ Glazman and K.\ A.\ Matveev, Zh.\ Eksp.\ Teor.\ Fiz.\
  {\bf 98}, 1834 (1990) [Sov.\ Phys.\ JETP {\bf 71}, 1031 (1990)].

\bibitem{Matveev2} K.~A.~Matveev, Zh.\ Eksp.\ Teor.\ Fiz.\ {\bf 99}, 1598
  (1991) [Sov.\ Phys.\ JETP {\bf 72}, 892 (1991)].

\bibitem{renorm-limit} The leading logarithm approximation
  (\ref{renorm-G}) is valid only as long as $\tilde G_{L,R} \ll G_0$.

\bibitem{NB} P.~Nozi\`eres and A.~Blandin, J.\ Phys.\ (Paris) {\bf 41},
  193 (1980); A.~I.~Larkin and V.~I.~Melnikov, Zh.\ Eksp.\ Teor.\ Fiz.\
  {\bf 61}, 1231 (1971) [Sov.\ Phys.\ JETP {\bf 34}, 656 (1972)].

\bibitem{note2} The detailed discussion on this reduction to effective
  one-dimensional models can be found in Ref.\ \cite{Matveev1}. A similar
  model has been used earlier by K.~Flensberg\cite{Flensberg}.

\bibitem{Sol} J.~S\'olyom, Adv.\ Phys.\ {\bf 28}, 201 (1979); V.~J.~Emery,
  in {\it Highly Conducting One-Dimensional Solids}, eds.\ J.\ Devreese
  {\it et al.} (Plenum, New York, 1979).

\bibitem{Flensberg} K.~Flensberg, Phys.\ Rev.\ B {\bf 48}, 11156 (1993).

\bibitem{Furusaki} A.~Furusaki and N.~Nagaosa, Phys.\ Rev.\ B {\bf 47},
  3827 (1993).

\bibitem{Kane} C.~L.~Kane and M.~P.~A.~Fisher, Phys.\ Rev.\ B {\bf 46},
  15233 (1992).

\bibitem{g=1/2} Since half of the degrees of freedom, the charging
  mode, is frozen due to the charging effect at $T<E_C$, the parameter
  $g$ becomes half of the original value $g=1$ \cite{Furusaki}.

\bibitem{Guinea} F.~Guinea, Phys.\ Rev.\ B {\bf 32}, 7518 (1985).

\bibitem{Fabrizio} M.~Fabrizio and A.~O.~Gogolin, Phys.\ Rev.\ B
  {\bf 50}, 17732 (1994).


\end{references}
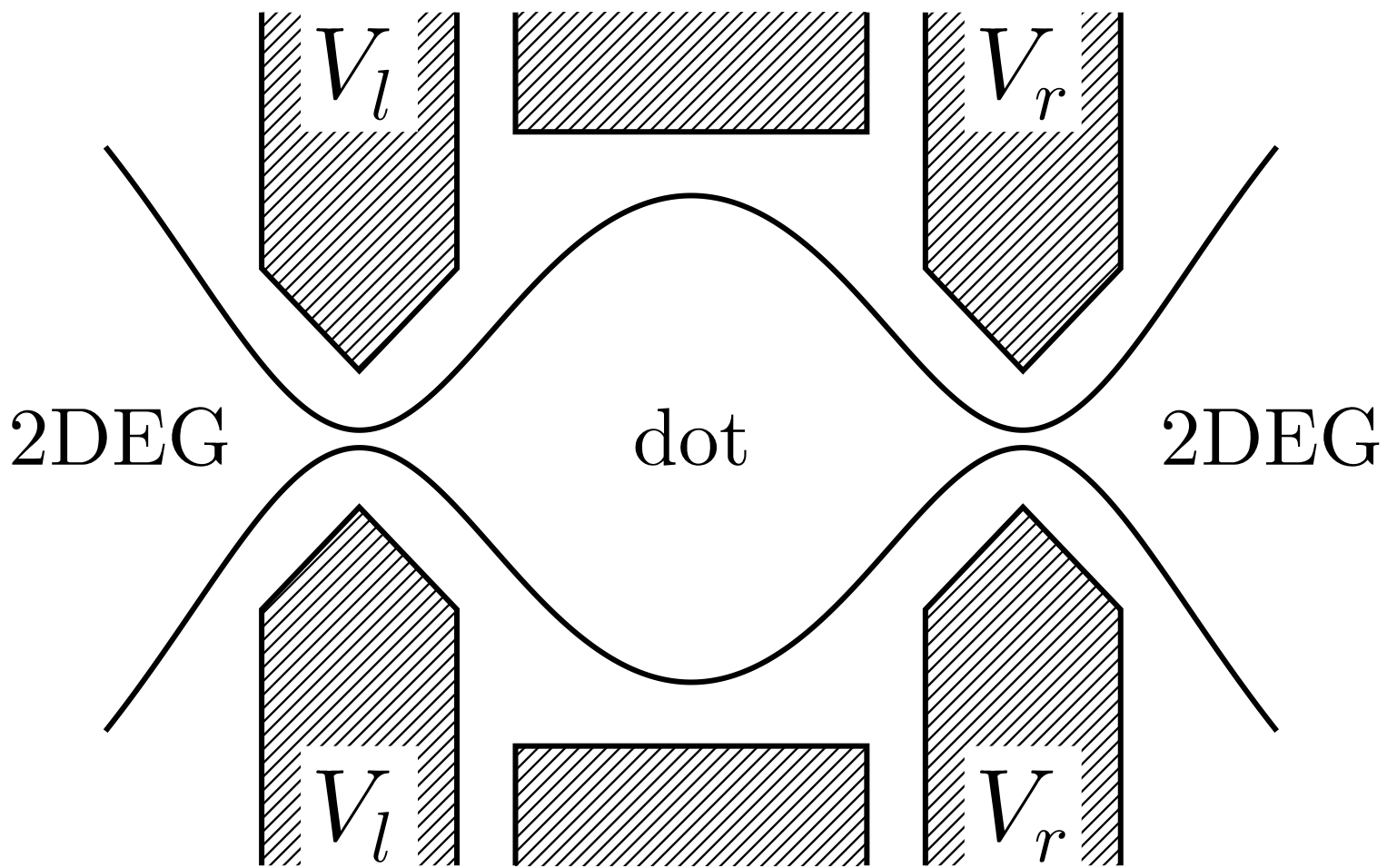
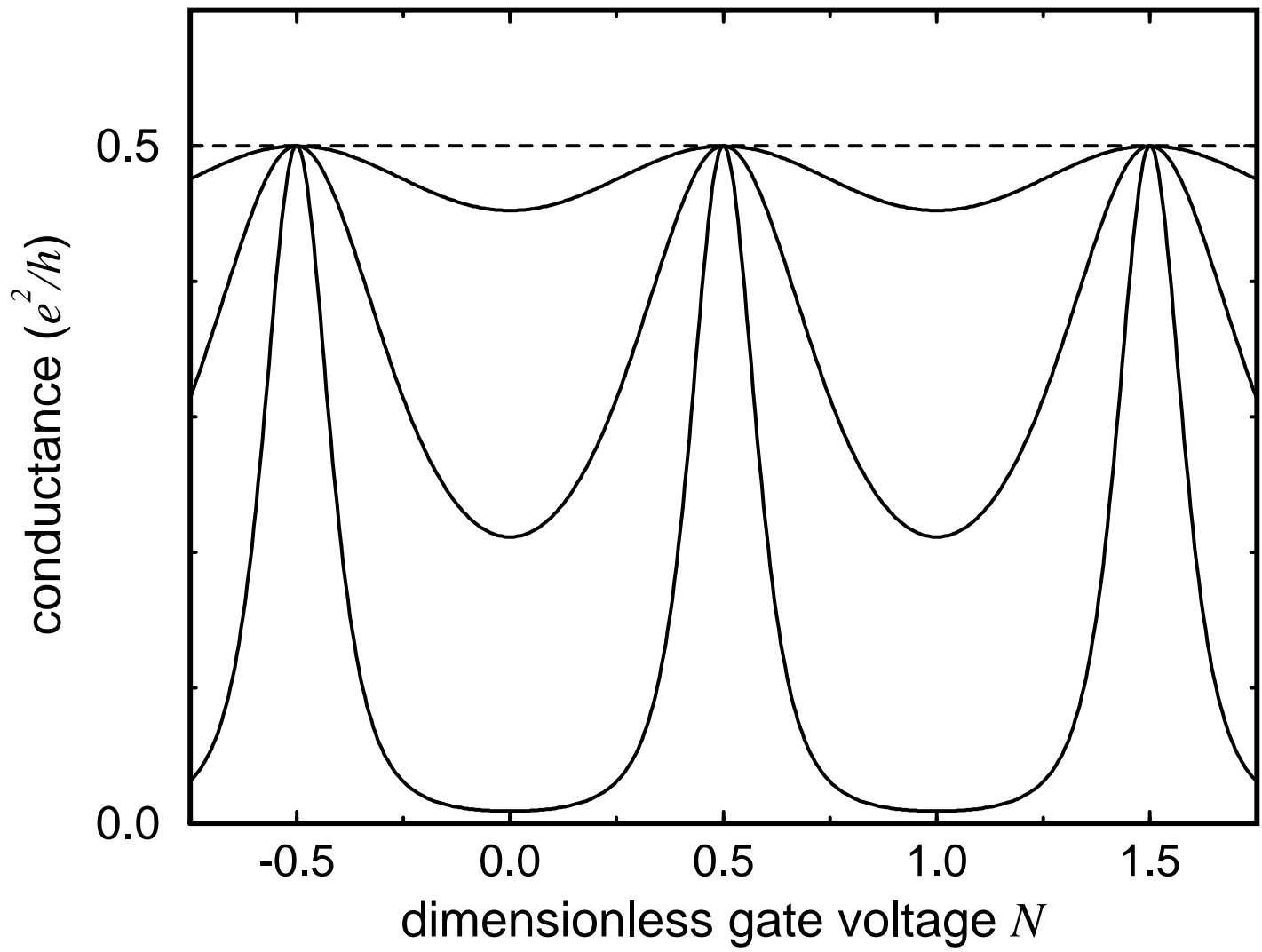
\end{document}